\documentclass[ALICE,manyauthors]{cernphprep}
\usepackage[comma,square,numbers,sort&compress]{natbib}
\usepackage{hyperref}
\usepackage{multirow}
\newcommand{\RV}{\mathrm{MB}_\mathrm{AND}}
\newcommand{\SV}{\sigma_\mathrm{MB_\mathrm{AND}}}
\newcommand{\RO}{\mathrm{MB}_\mathrm{OR}}
\newcommand{\SO}{\sigma_\mathrm{MB_\mathrm{OR}}}

\begin{document}%
%
%
\begin{titlepage}
\PHnumber{2013-082}                 
\PHdate{April 30, 2013}              
%
%
\title{Performance of the ALICE VZERO system}
\ShortTitle{Performance of the ALICE VZERO system}   
\Collaboration{ALICE Collaboration%
         \thanks{See Appendix~\ref{app:collab} for the list of collaboration
                      members}}
\ShortAuthor{ALICE Collaboration}      
\begin{abstract}
ALICE is an LHC experiment devoted to the study of strongly
interacting matter in proton--proton, proton--nucleus and nucleus--nucleus
collisions at ultra-relativistic energies. The ALICE VZERO system, made of two
scintillator arrays at asymmetric positions, one on each side of the interaction
point, plays a central role in ALICE. In addition to its core function as a
trigger source, 
the VZERO system is used to monitor LHC beam conditions, to reject beam-induced
backgrounds and to measure basic physics quantities such as luminosity, particle
multiplicity, centrality and event plane direction in nucleus-nucleus
collisions. 
After describing the VZERO system, this publication presents its performance
over more than four years of operation at the LHC.
\end{abstract}
\end{titlepage}
\setcounter{page}{2}

\section{Introduction}
\label{sec:intro}

ALICE (A Large Ion Collider Experiment) is an experiment dedicated to the study
of heavy-ion collisions 
at the LHC~\cite{ALICEPPRv1,ALICEPPRv2}. It is designed to explore the physics
of strongly interacting 
matter in proton--proton (pp) collisions and of Quark-Gluon Plasma (QGP) at
extreme values of energy density and 
temperature in nucleus--nucleus collisions. It provides identification of
particles up to the highest multiplicities 
and down to low transverse momentum
($p_\mathrm{T}\gtrsim$~50~MeV/$c$). The physics programme also includes 
proton--nucleus collisions to study the cold nuclear effects as well as lighter
ion collisions to vary energy density and 
interaction volume. Data taking in pp mode provides reference data for the
heavy-ion programme and allows to investigate 
various specific properties of the strong-interaction, thereby playing a
complementary role to that of the other LHC experiments.

The ALICE apparatus consists of a central barrel, a forward muon spectrometer
and a set of 
detectors in the forward regions including the VZERO system. This system
provides triggers for the 
experiment (minimum bias or centrality trigger) and separates beam--beam
interactions from background events 
such as beam--gas interactions, either at trigger level or in off-line analyses.
Furthermore, it is also used to 
measure beam luminosity, charged particle multiplicity and azimuthal
distributions. The control of the luminosity allows the
determination of the absolute cross-section value of reaction processes. The knowledge
of the charged particle multiplicity is 
essential for the evaluation of the centrality of nucleus--nucleus collisions.
The particle azimuthal 
distribution is important for determining the Pb--Pb collision reaction plane. 

In this publication, a description of the VZERO system and its Front-End
Electronics (Section~\ref{sec:V0arrays}) 
is given as well as the calibration procedure for individual channels
(Section~\ref{sec:calibration}). 
Details about the role of the system in the ALICE triggering scheme
(Section~\ref{sec:trigger}) are then described. 
Finally, its use for physics purposes, is also pointed out: monitoring of
colliding beams for the evaluation of luminosity 
(Section~\ref{sec:luminosity}); measurement of the multiplicity for the
estimation of the centrality of collisions 
(Section~\ref{sec:multiplicity}); measurement of azimuthal distributions of
charged particles for the knowledge of the 
anisotropic flow of collisions (Section~\ref{sec:reactionplane}). The last three
sections are only briefly covered, since detailed 
developments have already been published.


\section{The VZERO system}
\label{sec:V0arrays}

\begin{figure}[t]
\centering
\includegraphics[width=\columnwidth]{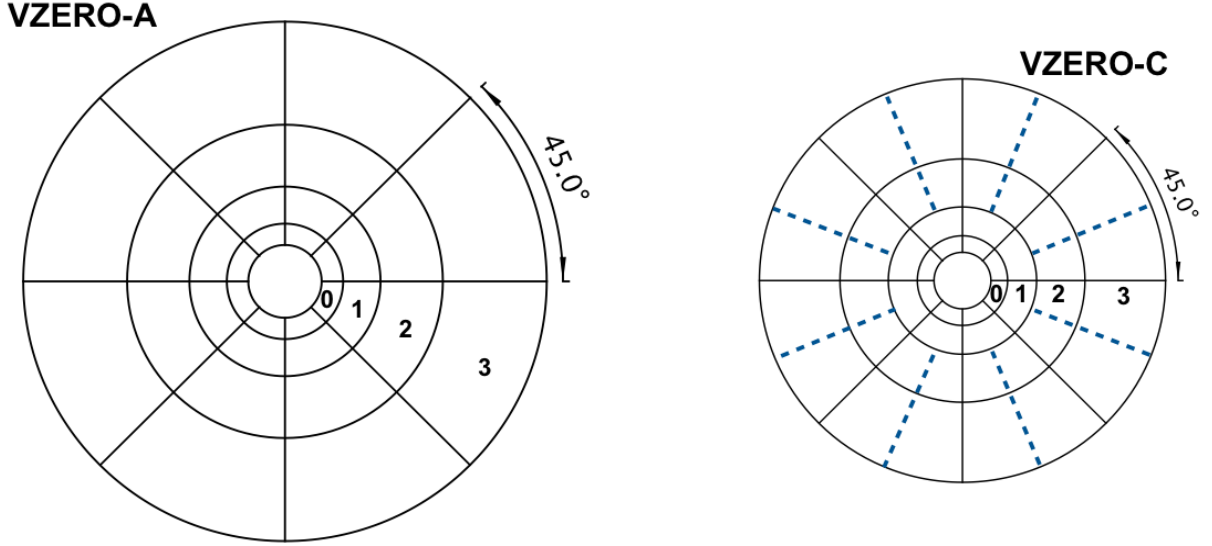}
\caption{Sketches of VZERO-A and VZERO-C arrays showing their segmentation.
Scintillator thicknesses are  2.5 and 2~cm respectively. Radii of rings are
given in Tab.~\protect\ref{tab:V0coverage}. The scintillator segments on both
sides of the dashed lines are connected to the same PMT (see
Section~\protect\ref{sec:calibration}).}
\label{fig:V0arrays}
\end{figure}

\begin{figure}
\centering
		\begin{minipage}[t]{\textwidth} 
			\begin{center}
			
\includegraphics[width=0.49\columnwidth]{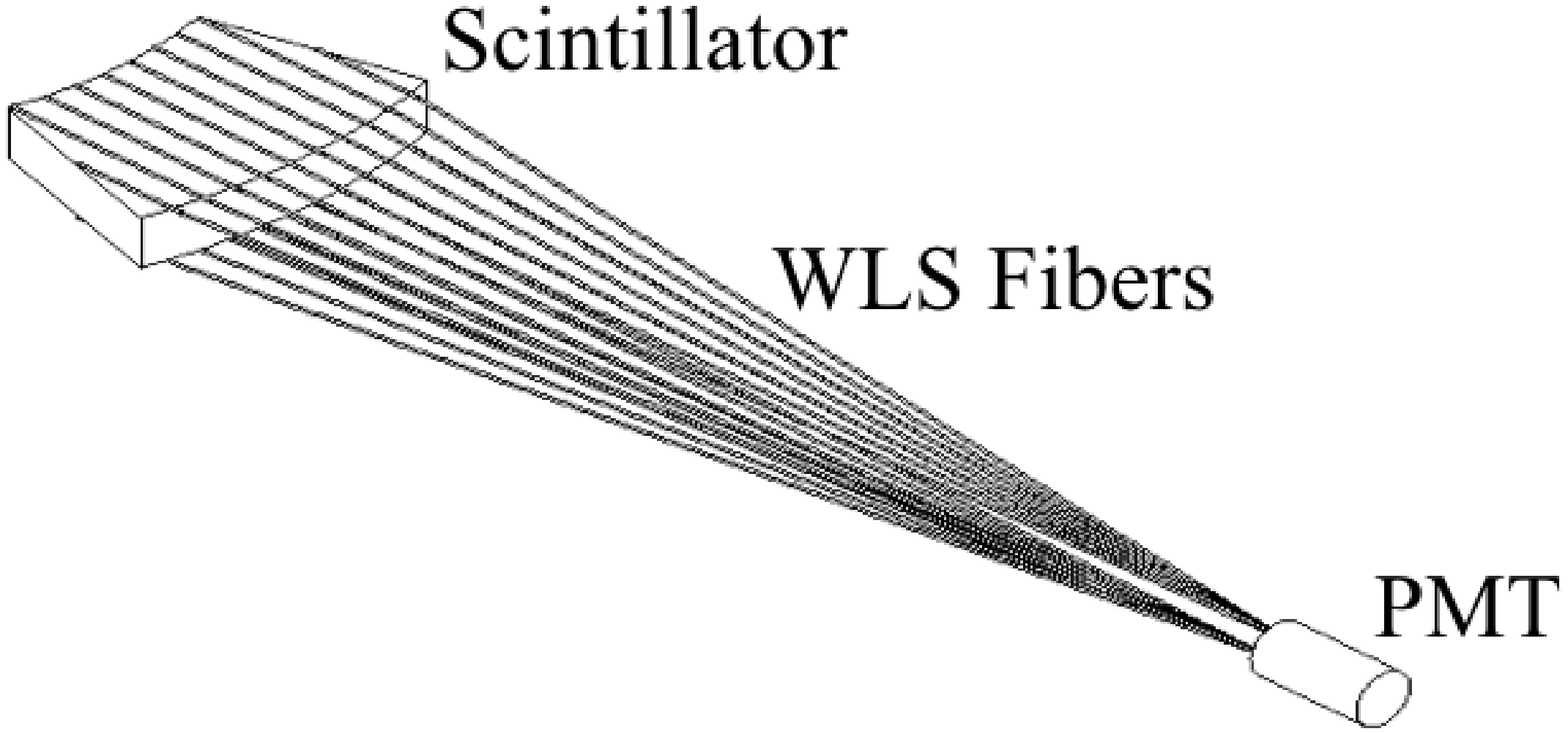}
				\hspace{0.cm}
			
\includegraphics[width=0.49\columnwidth]{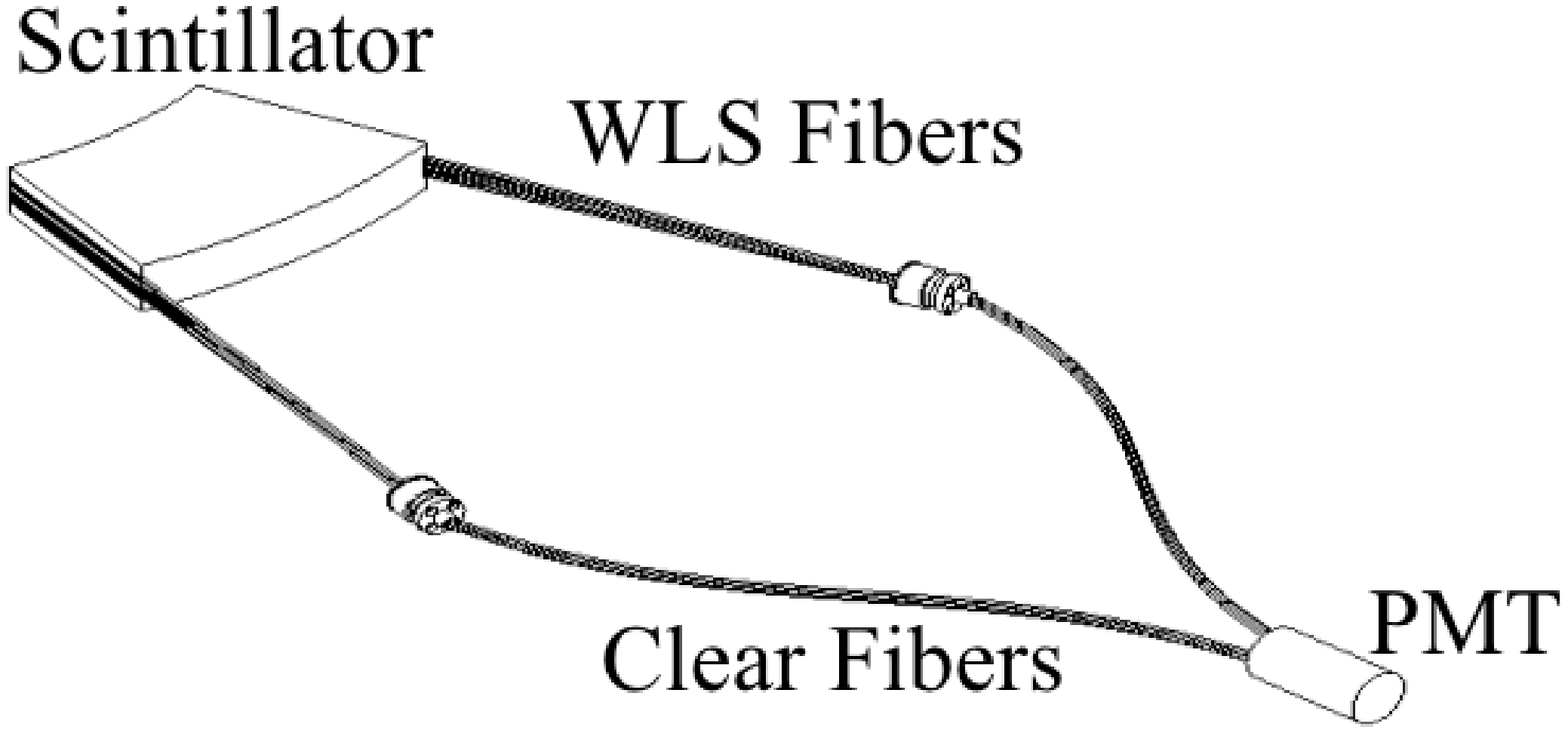}
			\end{center}
		\end{minipage}
\caption{Schematic drawings of elementary cell designs for VZERO-A (top) and
VZERO-C rings~0-1 (bottom). For VZERO-C rings~2-3, two scintillating sets
(scintillator and WLS fibers) are connected to a single PMT through four clear
fiber beams 
(see Fig.~\protect\ref{fig:V0arrays}).}
\label{fig:cellschematic}
\end{figure}

\begin{figure}
\centering
\includegraphics[width=\columnwidth]{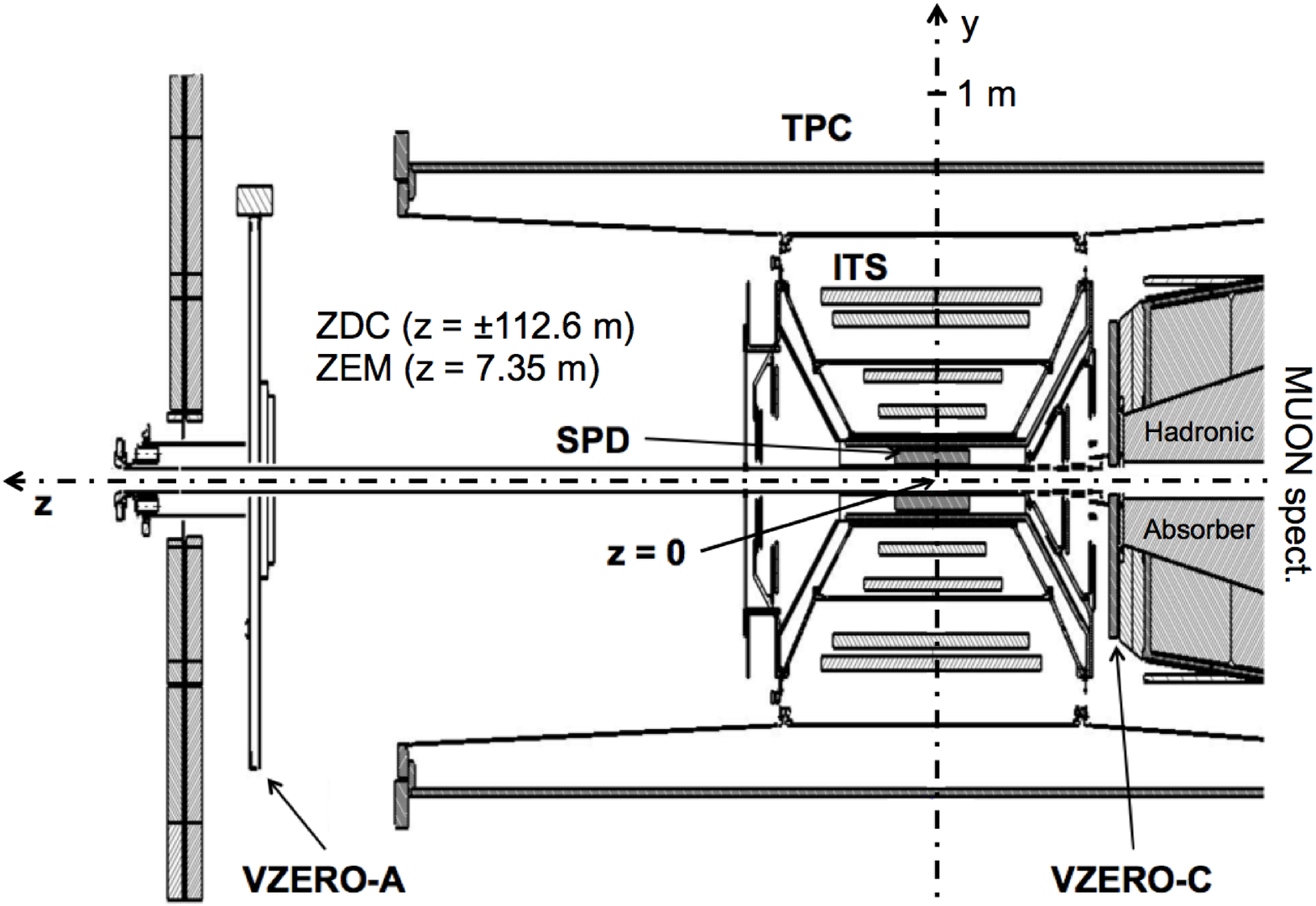}
\caption{Position of the two VZERO arrays, and of the few detectors quoted in
the text, 
within the general layout of the ALICE experiment.}
\label{fig:ALICEView}
\end{figure}

The detailed description of the VZERO system may be found
in~\cite{TDR,ALICEatLHC} and references therein. It is composed of two arrays,
VZERO-A and VZERO-C, which cover the pseudorapidity ranges $2.8<\eta<5.1$ and
$-3.7<\eta<-1.7$ for collisions at the nominal vertex ($z$~=~0). Each of the
VZERO arrays is segmented in four rings in the radial direction, and each ring
is divided in eight sections in the azimuthal direction
(Fig.~\ref{fig:V0arrays}). The pseudorapidity coverage of each ring is given in
Tab.~\ref{tab:V0coverage}. Each channel of both arrays is made of a BC404
plastic scintillator from Bicron~\cite{Bicron} with a thickness of 2.5 and
2.0~cm for VZERO-A and VZERO-C respectively. BCF9929A Wave-Length Shifting
(WLS) fibers from Bicron are embedded in both faces of the segments (VZERO-A) or
glued along their two radial edges (VZERO-C). Figure~\ref{fig:cellschematic}
shows a schematic view of an elementary cell of each array. These two different
designs were mandatory to comply with the integration constraints of each array.
The VZERO-A is located 329~cm from the nominal vertex ($z$~=~0) on the side
opposite to the muon spectrometer (Fig.~\ref{fig:ALICEView}). 
The VZERO-C is fixed on the front face of the hadronic absorber. The position of
the various rings of the VZERO-C along the $z$ direction is given in
Tab.~\ref{tab:V0coverage}.
The specific geometry of the VZERO-C array permits an optimized azimuthal
coverage while leaving room for the passage of the WLS fibers. As shown in
Fig.~\ref{fig:cellschematic} there are no such constraints for the VZERO-A
array. 

\begin{table}
\centering
\begin{tabular}{|c|c|c|c|c|c|c|c|c|}
  \hline
\multirow{2}{*}{Ring} & \multicolumn{4}{c|}{VZERO-A}
&\multicolumn{4}{c|}{VZERO-C} \\
  \cline{2-9}
    &$\eta_{\textrm{\scriptsize max}}$/$\eta_{\textrm{\scriptsize min}}$ &
$\theta_{\textrm{\scriptsize min}}$/$\theta_{\textrm{\scriptsize max}}$
&r$_{\textrm{\scriptsize min}}$/r$_{\textrm{\scriptsize max}}$& $z$ 
    &$\eta_{\textrm{\scriptsize min}}$/$\eta_{\textrm{\scriptsize max}}$&
$\theta_{\textrm{\scriptsize max}}$/$\theta_{\textrm{\scriptsize min}}$ &
r$_{\textrm{\scriptsize min}}$/r$_{\textrm{\scriptsize max}}$& $z$ \\
    \hline
   0 & 5.1/4.5 & 0.7/1.3 &  4.3/7.5 & 329 &-3.7/-3.2 & 177.0/175.3 &  4.5/7.1  &
-86\\
   1 & 4.5/3.9 & 1.3/2.3 &  7.7/13.7 & 329 &-3.2/-2.7 & 175.3/172.3 &  7.3/11.7 
& -87\\
   2 & 3.9/3.4 & 2.3/3.8 &  13.9/22.6 & 329 &-2.7/-2.2 & 172.3/167.6 & 
11.9/19.3 & -88\\
   3 & 3.4/2.8 & 3.8/6.9 &  22.8/41.2 & 329 &-2.2/-1.7 & 167.6/160.0 & 
19.5/32.0 & -88\\
  \hline
\end{tabular}
\vspace{2mm}
\caption{Pseudorapidity, angular acceptance (deg.), radius (cm) and $z$ (cm)
position along the beam axis
of VZERO-A and VZERO-C median plane rings, as seen from the origin of the
coordinate system.}
\label{tab:V0coverage}
\end{table}

The light is transferred to the R5946-70 photomultiplier tube (PMT)
(Hamamatsu~\cite{Hamamatsu}) directly for VZERO-A and through an extra 3~m of
PMMA (Poly(methyl methacrylate)) clear fibers (Mitsubishi~\cite{Mitsubishi}) for
VZERO-C (Fig.~\ref{fig:cellschematic}). Fine mesh PMTs were chosen for their
capacity to operate in a magnetic field, since the VZERO arrays and their PMTs
are inside the ALICE solenoid which produces a field of 0.5~T. In order to
minimize the signal attenuation, the tube axis are oriented at 30$^\circ$
relatively to the magnetic field direction. For the two outer rings of the
VZERO-C array, each sector is divided into two segments of 22.5$^\circ$
(Fig.~\ref{fig:V0arrays}) both connected to the same PMT. This separation was
necessary to optimize the light collection by the WLS fibers glued on the radial
edges of the segments. 

For each elementary cell, the PMT output is split into two signals sent to the
Front-End Electronics (FEE), one of them after amplification by a factor 10. For
the 32 channels of each array~\cite{V0FEE}, the pulse time (leading time)
relative to the 40~MHz LHC bunch clock and the width at discriminator threshold
are measured by time to digital converters (TDC) on amplified signals, and the
charge is integrated by charge analog to digital converters (ADC) on direct signals. Two
types of trigger algorithms are implemented independently for VZERO-A and
VZERO-C arrays. The first trigger algorithm type, based on pre-adjusted time windows
corresponding to beam-beam or beam-background in coincidence with the time
signals from the counters, is used to select minimum bias events and to reject beam-induced background events. 
The second trigger algorithm type is based on the total charge collected by
each array, out of which events corresponding to Pb--Pb collisions
of centralities 0--10\% to 0--50\% are selected. Finally, five trigger sources~\cite{V0FEE}, chosen
according to specific features of collision mode and recorded data, are sent to
the Central Trigger Processor (CTP)~\cite{TriggerTDR}. 
The CTP collects signals from all trigger devices, provides trigger decisions
and distributes global trigger signals to the whole ALICE experiment, together
with the LHC clock and timing synchronization information. ALICE has a
multi-level trigger protocol, L0, L1 and L2, with latency with respect to the
collision time of 1.2, 6.5 and 88~$\mu$s, respectively. The VZERO arrays
contribute to the L0 trigger decision. 


\section{Detector calibrations and data corrections}
\label{sec:calibration}

\begin{figure}
\centering
		\begin{minipage}[t]{\textwidth} 
			\begin{center}
			   
\includegraphics[width=0.49\columnwidth]{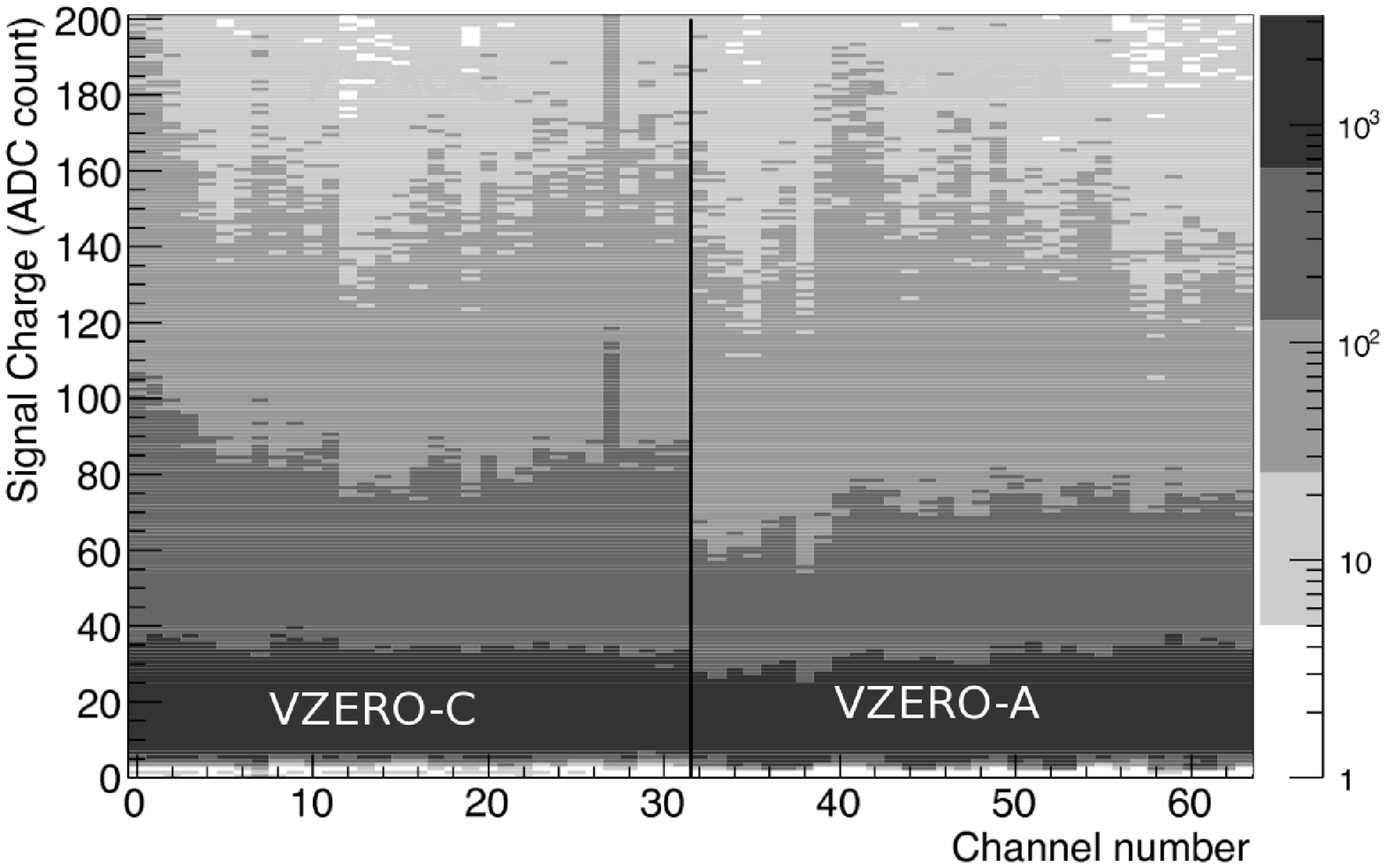}
			    \hspace{0.cm}
			
\includegraphics[width=0.49\columnwidth]{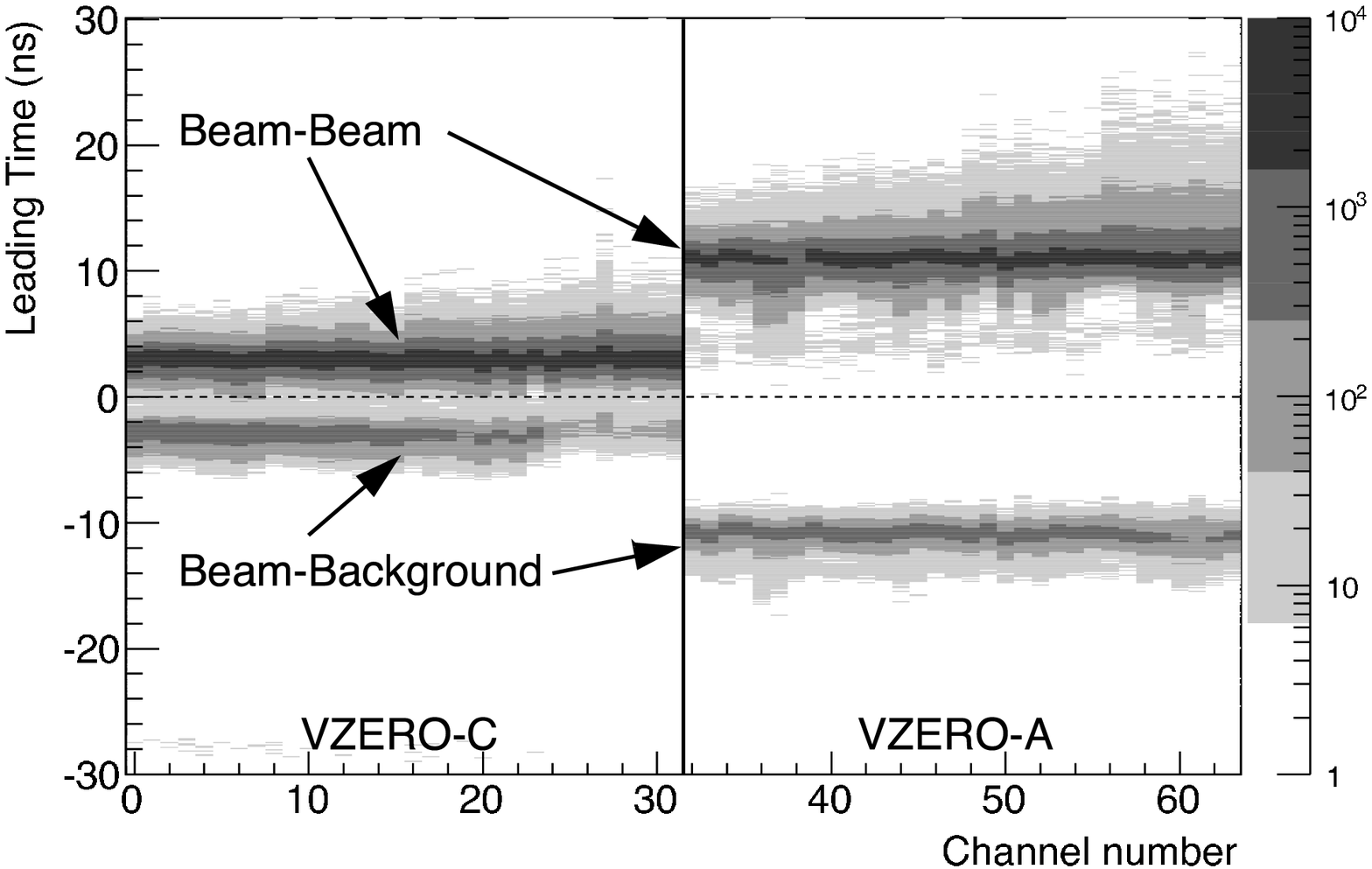}
			\end{center}
		\end{minipage}
\caption{Charge of the pulse in ADC counts (top) and leading time of the pulse
in nanoseconds (bottom)
versus channel number for pp collisions at $\sqrt{s}=$~7~TeV. Channels numbered
from 0 to 
31 correspond to VZERO-C, channels numbered from 32 to 63 correspond to
VZERO-A.}
\label{fig:rawtimeadc}
\end{figure}

For each channel, two quantities need to be calibrated to ensure uniform output
for all the channels of the 
detector: the PMT gain and the light yield of the scintillator-fibers assembly.
The latter takes 
into account both the light yield of the scintillator elements and the light
transmission through the 
WLS fibers and optical fiber lines. The calibration of the detector was carried
out in two steps. The gain curve of each PMT was determined in the laboratory
using cosmic muons passing through a scintillator cell. Measurements at three
high voltages were carried out. The high charge resolution of the XP2020 PMT
(Photonis~\cite{XP2020}) calibrated with a LED source, made it possible to
measure the average number of photo-electrons produced in the VZERO
scintillators, thus providing an absolute gain measurement of VZERO PMTs. The
light yield of each optical line was measured in situ with pp collisions at
$\sqrt{s}$~=~7~TeV. Using a simulation of the detector and knowing the gain vs.
high voltage curve of each PMT, it was possible to determine the light yields of
individual channels. The high voltage of each channel was then adjusted to have
the same response to the passage of one minimum ionizing particle (MIP) in terms
of ADC counts (about 15 ADC counts per MIP in pp collision mode). 
The resulting amplitude distributions are uniform within 10--20\%
(Fig.~\ref{fig:rawtimeadc}-top). The average high voltage in pp collision mode
is around 2000~V. To cope with the large charged particle multiplicity in Pb--Pb
collisions without saturating the ADCs, the voltages are reduced to about
1500~V, corresponding to a reduction factor of about 10 for the PMTs gains in
comparison with pp voltages.

Relative time adjustment between channels was done using a cable with a length
known with a precision of 1~ns. Fine tuning adjustment was performed using
individual programmable delays on the FEE boards, resulting in a uniform
response of the elementary cells (Fig.~\ref{fig:rawtimeadc}-bottom). The time of
flight of particles coming from the Interaction Point (IP) to the VZERO-A
(VZERO-C) is about 11~ns (3~ns). The dashed line in
Fig.~\ref{fig:rawtimeadc}-bottom represents the collision time. For beam induced
background coming from the C~side (A~side) to the VZERO-A (VZERO-C) the time of
flight is also 11~ns (3~ns). Furthermore, a signal arriving at -11~ns (-3~ns) to
the VZERO-A (VZERO-C) can be clearly seen. It corresponds to particles from beam
induced background coming from the A side (C~side). In the case of backgrounds,
the dashed line represents the time when particles cross the vertical plane at
$z$~=~0.

\begin{figure}
\centering
\includegraphics[width=0.9\columnwidth]{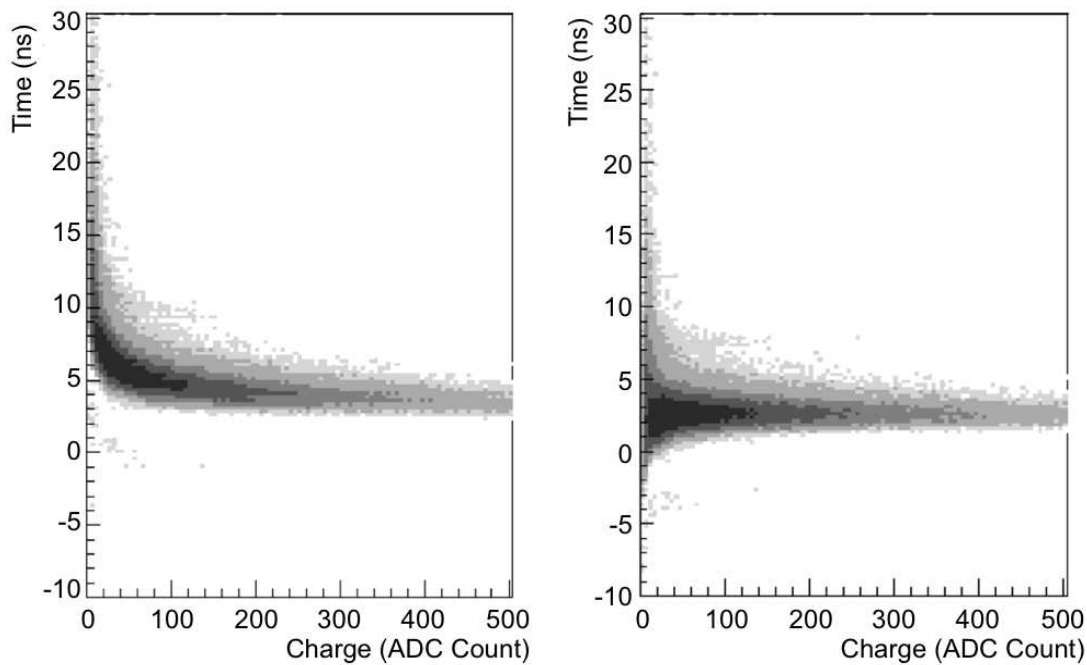}
\caption{Correlation between leading time in nanoseconds and charge of the PMT
signals in ADC channel 
counts for Pb--Pb collisions, before any correction (left) and after slewing
correction (right).}
\label{fig:slewingtimeadc}
\end{figure}

Due to the fact that threshold discriminators are used, the leading time
measurement is affected by a slewing effect. The measured timing of the rising signal edge
depends on the amplitude as shown in Fig.~\ref{fig:slewingtimeadc}-left. 
The measured raw time $t_{\textrm{\scriptsize raw}}$ is corrected for slewing effect by subtracting an offset $\Delta t(Q)$, function of the pulse charge $Q$: $t_{\textrm{\scriptsize corr}} = t_{\textrm{\scriptsize raw}} - \Delta t (Q)$. 
The offset calculation uses the following parametrization:
\begin{equation}
\Delta t (Q) = C \sqrt{\frac{s}{Q}},
\label{eq:Slewing}
\end{equation}
where $C$ is a constant equal to 10.5 ns and $s$ the threshold setting of the
discriminator which is fixed to 
1 (4) ADC channel(s) for Pb--Pb (pp) collisions.

The effect of this correction is shown in Fig.~\ref{fig:slewingtimeadc}-right.
The average time response is then independent of the charge delivered by the
PMT. The time resolution dependence on the signal can be deduced from this plot.
Averaged over all signal amplitudes, the individual channel time resolution is
of the order of 1~ns for both arrays. This time resolution is independent of the
colliding system (pp or Pb--Pb).

Since 2009, the VZERO detector has been exposed to radiation produced by the
LHC. Both beam--beam collisions and beam-induced backgrounds 
contribute to the accumulated radiation dose. We observed a loss in signal
amplitude of 10 to 40\% at the end of 2012, depending on the channel, which is
attributed to ageing effect. An accurate determination of the accumulated dose
is difficult due to uncertainties in the beam-induced background evaluation.
Nevertheless, from the luminosity measurement and the background monitoring, we
can estimate that the total accumulated dose per channel does not exceed
50~krad. This value is six times lower than the maximum dose accumulated during
a radiation tolerance test using a 60~MeV proton beam~\cite{radiations}. Results
from this test show that the loss of signal from the scintillating material is
in the order of a few percent at 50~krad, which cannot explain the observed
reduction of signal amplitude. The dominant contribution to the loss is
attributed to ageing of the PMT photocathodes. This effect was shown by
monitoring the signal amplitude from direct particle hits on the PMT compared to
the signal amplitude from scintillation, the discrimination between both
components being possible as the timing difference between these two sources is
around 6~ns for the VZERO-A side and even longer for the VZERO-C side. 
Early 2013, after the LHC shutdown, a set of 6 PMT chosen randomly were extracted from the experiment to be tested in laboratory.
Their gains measured in 2013, compared to the one measured in 2006 before the installation, show the same losses than the ones measured in the experiment.
To compensate for the signal loss and adjust the gain approximately at its initial
value, the individual PMT high voltage had to be increased accordingly.


\section{The VZERO trigger system}
\label{sec:trigger}

One of the primary roles of the VZERO system is to provide the ALICE experiment
with a minimum bias (MB) trigger both in pp and Pb--Pb collisions and with
centrality based triggers in Pb--Pb mode. The VZERO system has been operational
throughout the whole period of ALICE operating since 2009, thus demonstrating
the stability and robustness of the detector and its readout system.

In the 2009 and 2010 pp data taking periods, the ALICE MB trigger $\RO$ was built
requiring a hit in the Silicon Pixel Detector (SPD in Fig.~\ref{fig:ALICEView})
which is the innermost part of the Inner Tracking System
(ITS)~\cite{ALICEatLHC} 
or in either of the VZERO arrays (VZERO-A or VZERO-C). The threshold setting
of the VZERO individual channels corresponds approximately to one quarter of the
mean energy deposit by a minimum ionizing particle. This MB trigger was set in
coincidence with signals from two beam pick-up counters, one on each side of the
interaction region, indicating the passage of proton bunches. This trigger
corresponds to the requirement of at least one charged particle anywhere in
8~units of pseudorapidity. During the 2011 and 2012 pp data taking periods, with
the increasing LHC luminosity and beam background, the trigger moved to a
stricter condition using the coincidence either between both VZERO arrays and
the LHC bunch crossing signals, or between both VZERO arrays, the LHC bunch
crossing signals and any other detector triggering on specific event topology like the muon spectrometer or the electromagnetic calorimeter. 
The last configuration was used to search for rare physics signals. 

\begin{figure}[t]
\centering
\includegraphics[width=0.7\columnwidth]{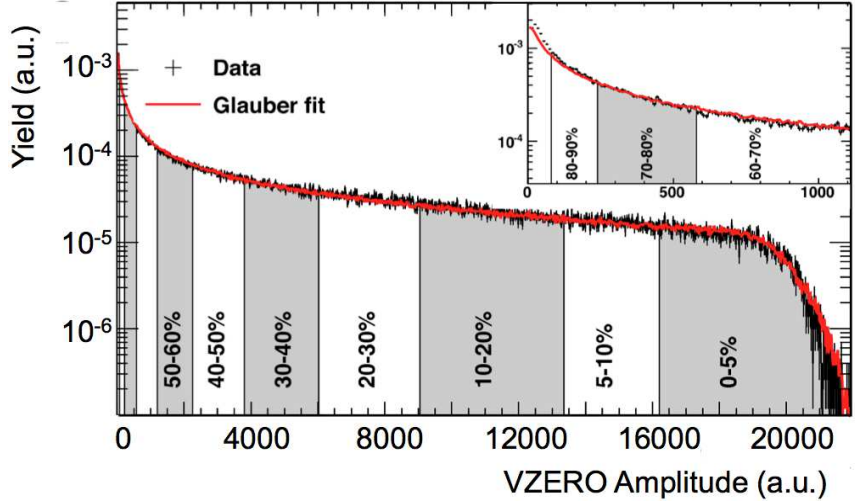}
\caption{Distribution of the sum of amplitudes in the two VZERO arrays (black
histogram) in Pb--Pb 
collisions at $\sqrt{s_{\rm NN}}$~=~2.76 TeV~\cite{CentralityPaper1}. The red
line shows the fit with a Glauber 
model~\cite{Miller}. The shaded areas define the different centrality classes of
hadronic collisions. 
The inset shows the low amplitude part of the distribution.}
\label{fig:Centrality}
\end{figure}

In Pb--Pb collisions, the interaction trigger is configured to obtain high
efficiency for hadronic interactions, requiring at least two out of the
following three conditions: (i) two pixel chips hit in the outer layer of the
SPD, (ii) a signal in VZERO-A, (iii) a signal in VZERO-C. The threshold setting
corresponds approximately to the mean energy deposited by one minimum ionizing
particle. The PMT gains are reduced by a factor of 10 in order not to saturate
the ADC for the most central collisions. At the end of the 2010 Pb--Pb  data
taking period, as the beam luminosity increased, the minimum bias trigger had to
be restricted to the coincidence between SPD, VZERO-A and VZERO-C in order to
suppress signals created by electromagnetic dissociation of Pb nuclei as much as
possible. During the full 2011 Pb--Pb run, the VZERO delivered three different
triggers: (i) the VZERO-A and VZERO-C coincidence asking for at least one cell
hit in each array, (ii) a trigger signal selecting the collisions corresponding to a centrality of 0--50\%, (iii) a
trigger signal selecting the  0--10\% most central collisions (see Fig.~\ref{fig:Centrality}).

In both running modes (pp and Pb--Pb), a non-negligible background comes from
interactions between the beams and the residual gas within the beam pipe and
from interactions between the beam halo and various components of the
accelerator such as collimators. Using the time of flight of particles detected
by each VZERO array, particles coming from collisions and particles coming from
beam-gas background can be clearly distinguished. 

\begin{figure}[t]
\centering
		\begin{minipage}[t]{\textwidth} 
			\begin{center}
			   
\includegraphics[width=0.47\columnwidth]{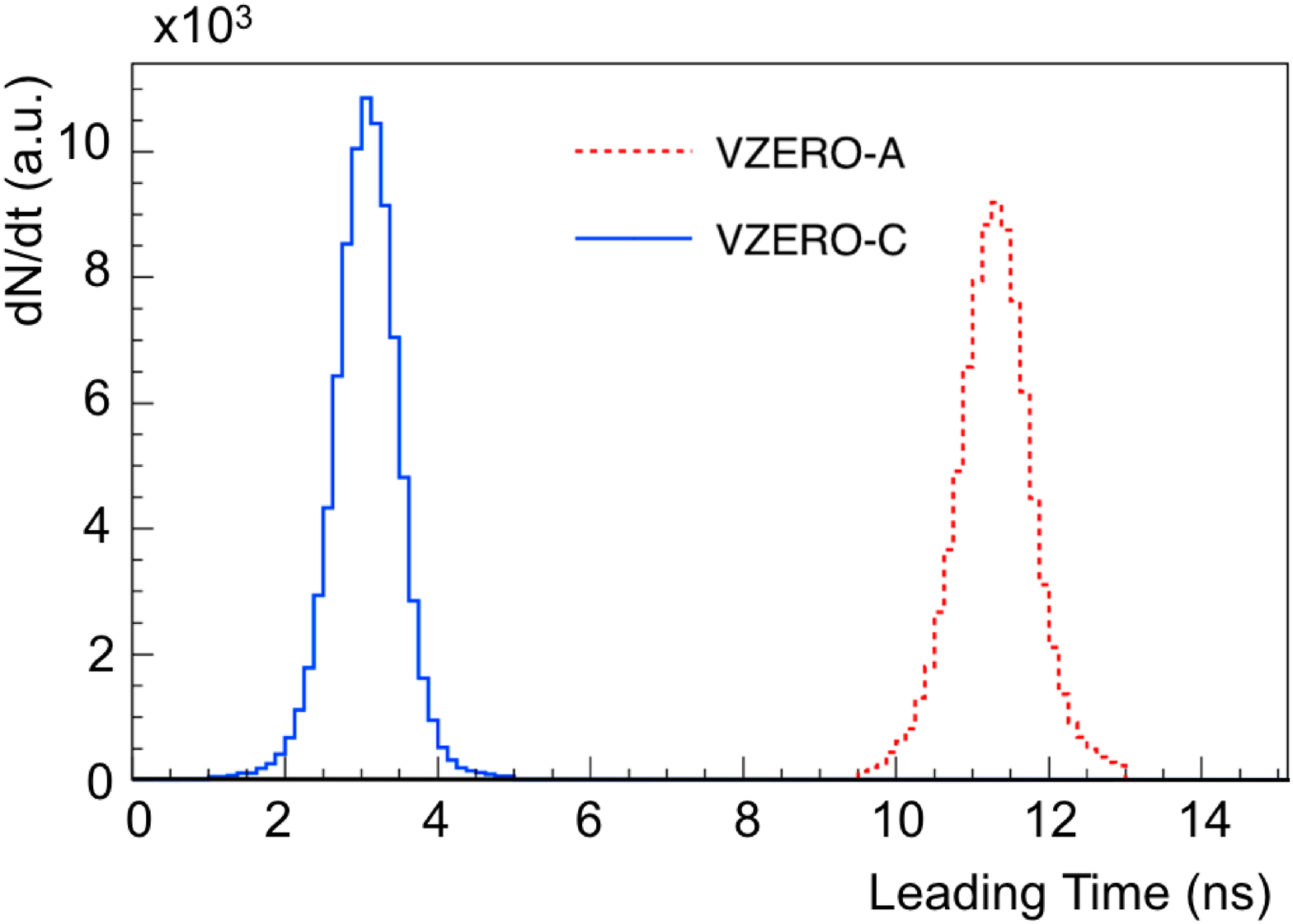}
			    \hspace{0.cm}
			
\includegraphics[width=0.49\columnwidth]{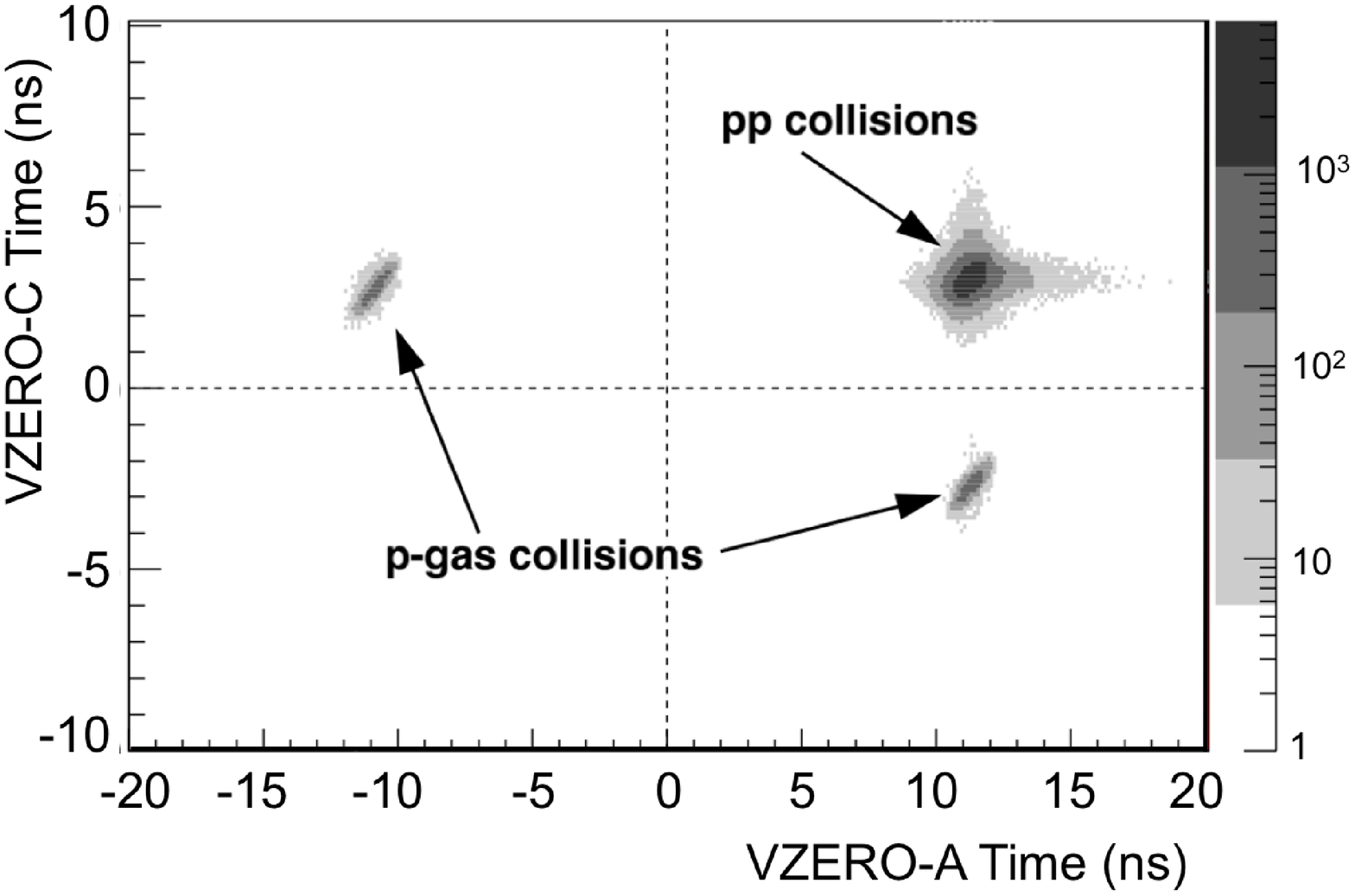}
			\end{center}
		\end{minipage}
\caption{Top: VZERO-A (red-dashed line) and VZERO-C (blue-solid line) weighted
average time of flight distributions for beam-beam collisions, with 0.45 and
0.35~ns~r.m.s. respectively. Bottom: weighted average time of flight (as defined
by 
Eq.~\protect \ref{eq:timeaverage}) of the particles detected in VZERO-C versus
VZERO-A. The dashed line intersection represents the time of the collisions at
the interaction point, or the crossing time of the background tracks at the
vertical plane $z$~=~0.}
\label{fig:V0A_V0C_Time}
\end{figure}

Two beam-gas background selections can be applied thanks to the VZERO system, one online at the FEE level and one offline at the data reconstruction stage.
In the FEE, coincidence windows of 8~ns in length are placed around the beam-beam timing in order to select the beam-beam events and reject most of the beam-induced background events. It increases the fraction of good events recorded in the data. 
This information is also used in order to monitor the level of beam induced background.
This selection was in production starting with the 2012 run due to the increase of luminosity and beam-induced background level. 

The second selection, performed offline, is more refined.
For each VZERO array, a
weighted average time of flight over the channels above threshold is calculated.
As shown in Fig.~\ref{fig:slewingtimeadc}-right, the time measurement resolution
is better for larger amplitude signals. Therefore, for each channel hit, a
weight function $\omega(Q) = 1/ \sigma^2(Q)$ is calculated where $\sigma(Q)$ is
the channel time resolution parametrized as:
\begin{equation}
\sigma(Q) = \sqrt{a^2 + \frac{b^2}{N_{e}} +  c^2  \frac{s}{Q^3}},
\label{eq:weight}
\end{equation}
where $N_{e}$ is the mean number of photo-electrons obtained from the
pulse charge and the PMT gain (see Section~\ref{sec:calibration}), $s$ the
threshold setting of the discriminator (in ADC channel) and $a$ (0.39~ns), $b$
(2.5~ns) and $c$ (15.8~ns*ADC~channel) the parameters extracted from a fit of
the time distributions after slewing correction
(Fig.~\ref{fig:slewingtimeadc}-right). The first term of this equation is
related to the intrinsic resolution of the detector and the second term to the
photo-electron statistics. The last term comes from the uncertainty of the
slewing correction due to 
fluctuations of the collected charge (derivative of Eq.~\ref{eq:Slewing}).

The weighted average time of flight of each array is then calculated as:
\begin{equation}
\label{eq:timeaverage}
\langle T \rangle = \frac{\sum_{i=1}^{32} \omega_i(Q_i) \times
T_i}{\sum_{i=1}^{32} \omega_i(Q_i)},
\end{equation}
where $T_i$ and $\omega_i(Q_i)$ are respectively the leading time and the weight
(calculated using 
Eq.~\ref{eq:weight}) for channel $i$. One average time is calculated for each
array, and the sum 
runs over all fired channels of the corresponding array. Time resolutions of
about 450~ps and 350~ps 
are achieved for VZERO-A and VZERO-C arrays (Fig.~\ref{fig:V0A_V0C_Time}-top).

Finally, the VZERO system allows rejection of events coming from beam-induced
backgrounds, by requiring a positive 
time of flight for both arrays (Fig.~\ref{fig:V0A_V0C_Time}-bottom). 

\begin{figure}
\centering
\includegraphics[width=0.5\columnwidth]{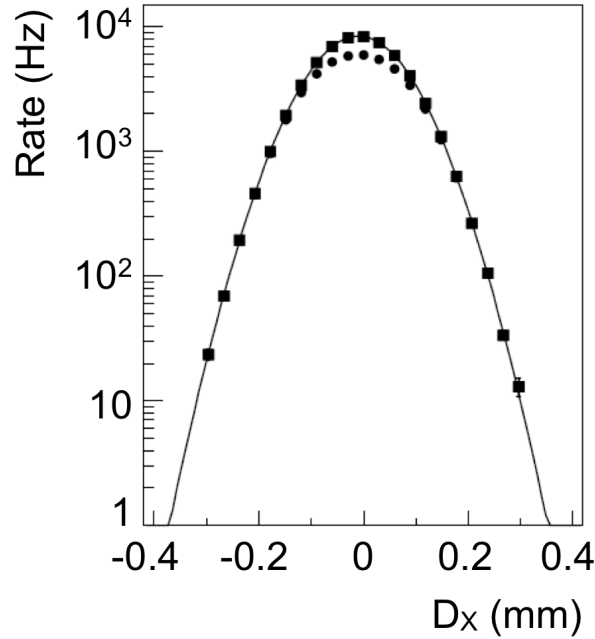}
\caption{Example of $\RV$ counting rate as a function of the beam displacement
in the horizontal direction 
from the van der Meer scan data taken in May 2010 at an energy of 7~TeV in pp
collisions. Dots represents the raw 
trigger rates, and squares represents the interaction rates after corrections (see
reference~\cite{CrossSection}). 
The line is used to guide the eye. The integral is calculated using the sum of
all bins.}
\label{fig:VdMScan}
\end{figure}

\section{Luminosity measurement}
\label{sec:luminosity}

The measurement of the absolute value of a reference process cross-section
allows the determination of an absolute normalization scale for other
cross-section measurements in the experiment~\cite{BCNWG}. In ALICE, the on-line
monitoring of the luminosity uses a time coincidence between the two VZERO
arrays. The rate $R$ of this coincidence is given by~\cite{VDMOYAMA}
\begin{equation}
\label{eq:Lumi1}
R =A\cdot\epsilon\cdot\sigma\cdot\mathcal{L}.
\end{equation}
where $\sigma$ is the inelastic cross-section, $A$ the acceptance, $\epsilon$
the efficiency and $\mathcal{L}$ the luminosity.

A measurement of the luminosity was carried out using the van der Meer scan
method~\cite{VdM} which measures 
the size and the shape of the colliding beams by observing the counting rate $R$
in a suitable 
monitor system while sweeping the two beams through each other. The rate
corresponding to the coincidence between VZERO-A 
and VZERO-C signals is named $\RV$. The luminosity $\mathcal{L}$ and therefore
the rate of $\RV(D_x,D_y)$ are functions of the transverse displacements $D_x$ and
$D_y$ of the beams. 

The shape factors $Q_x$ and $Q_y$ are obtained from the ratio of the area to the
height of the rate curve 
in both directions (Fig.~\ref{fig:VdMScan} for the $x$ direction). For a zero
degree beam crossing angle of beams with intensities $N_1$ and $N_2$ measured by
the LHC~\cite{BCNWG}, the luminosity $\mathcal{L}$, and hence the cross-section
$\SV$ corresponding to the rate of $\RV$, are obtained by:
\begin{equation}
  \mathcal{L}=k_b f N_1 N_2 Q_x Q_y \;\;\; \mbox{and} \;\;\; \SV= \RV(0,0) /
\mathcal{L}  
\label{Eq:LumiandCS},
\end{equation}
where $k_b$ is the number of colliding bunches and $f=11.2455$~kHz the LHC
revolution frequency.

The cross-section $\SV$ was measured for both pp colliding energies (2.76~TeV
and 7~TeV) to be 
$\SV (2.76~\mathrm{TeV})= 47.7 \pm 0.9 ~\mathrm{(syst.)} ~\mathrm{mb}$ and 
$\SV (7~\mathrm{TeV})= 54.3 \pm 1.9 ~\mathrm{(syst.)} ~\mathrm{mb}$. As a
coincidence between VZERO-A and VZERO-C signals is required, this cross-section
is not the minimum bias cross-section $\SO$. The triggering efficiency
($\SV/\SO$) was obtained from adjusted Monte Carlo simulations and evaluated to
(76.0 $^{+5.2}_{-2.8}$)\% and (74.2 $^{+5.0}_{-2.0}$)\% respectively. A detailed
description of this analysis can be found in reference~\cite{CrossSection}. This
cross-section is then used between successive van der Meer scans to monitor the
instantaneous luminosity using the $\RV$ rate measurement. Hence, the VZERO
system provides the integrated luminosity measurement for the ALICE experiment.

\begin{figure}
\centering
\includegraphics[width=0.7\columnwidth]{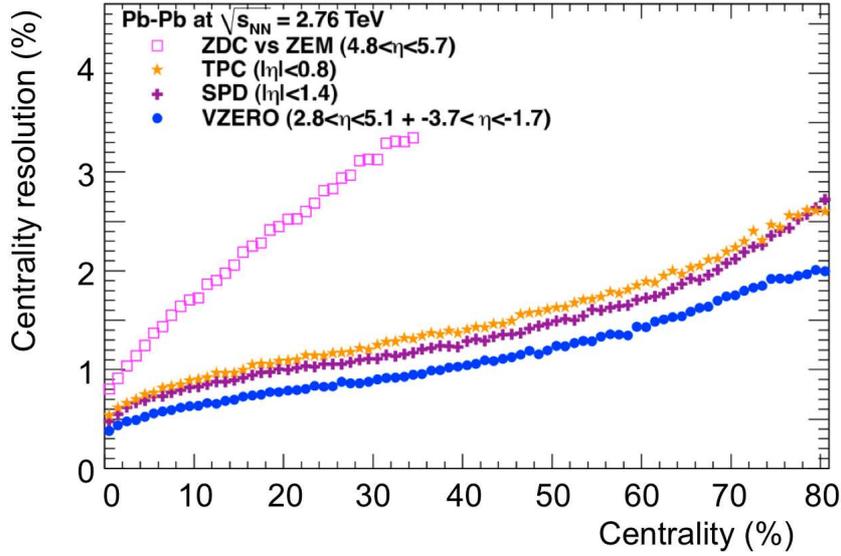}
\caption{Centrality percentile resolution versus centrality for various ALICE
detectors (see Fig.~\protect\ref{fig:ALICEView}). 
This figure is extracted from~\cite{CentralityPaper3}.}
\label{fig:CentralityResolution}
\end{figure}

\section{Multiplicity and centrality measurements}
\label{sec:multiplicity}

The VZERO system also provides a charged particle multiplicity measurement based
on the 
energy deposited in the scintillators. Using a detailed simulation of the
apparatus, the relation between the total charge collected inside a VZERO ring
and the number of primary charged particles emitted into the corresponding
pseudorapidity range was extracted.
Therefore, it was possible to obtain the charged particle multiplicity,
$\textrm{d}N_\textrm{\scriptsize ch}/\textrm{d}\eta$, in eight pseudorapidity
bins corresponding to the coverage of the different rings of the VZERO. This
coverage was increased during the 2010 Pb--Pb run, for which satellite
collisions at displaced vertices provided data~\cite{MultiplicityPaper,CentralityPaper1,CentralityPaper2}. These
displaced vertices are found in the range $-187.5<z<375$~cm and were due to ions
being captured in neighboring RF buckets during the beam injection procedure.
The resulting effect is an increase of the $\eta$ range covered by each array.
In particular the $\eta$ range seen by the VZERO-A array was extended to
$-2.81<\eta<5.22$.

Figure~\ref{fig:Centrality} represents a typical distribution of the VZERO
amplitudes~\cite{CentralityPaper1}. The shaded areas correspond to the different
centrality classes used in the physics analysis. The distribution is fit using a
Glauber model~\cite{Miller} which reproduces correctly the VZERO amplitude
distribution, hence the centrality, down to very low amplitudes (corresponding
to peripheral events) where contribution from electromagnetic interactions
between two lead ions is dominating (inset in Fig.~\ref{fig:Centrality}).
Figure~\ref{fig:CentralityResolution} shows the centrality percentile resolution
as a function of centrality for ZDC vs ZEM, TPC, SPD and VZERO
systems~\cite{CentralityPaper3}. The VZERO is used as the default in the ALICE
experiment as it gives the best resolution over the inspected centrality range.


\section{Event plane determination}
\label{sec:reactionplane}

Thanks to its granularity in the azimuthal direction (8 sectors of 45$^{\circ}$,
see Fig.~\ref{fig:V0arrays}), the VZERO system can be used to get an
experimental estimate of the reaction plane, defined by the beam direction and
the line between the centers of the two nuclei along which the impact parameter
is measured. The event plane is a key element in the study of the anisotropic
particle flow which develops relative to that plane and reflects properties of
the system in the early stages of the collision. In practice, the anisotropic
flow~\cite{Selyuzhenkov} is characterized with the help of a Fourier expansion
of the Lorentz invariant azimuthal distribution of  particles:
\begin{equation}
\rho(\phi-\Psi_\mathrm{RP}) = \frac{1}{2\pi}\left(1+2\sum_{n=1}^\infty v_n 
\cos[n(\phi-\Psi_\mathrm{RP})]\right),
\label{eq:fourier}
\end{equation}
where $\phi$ is the particle azimuthal angle, $\Psi_\mathrm{RP}$ the reaction
plane angle as defined above 
and $v_n$ the $n$-th harmonic of the anisotropic flow. Experimentally, the true
reaction plane can only
be reconstructed approximately due to interactions within the produced matter. In
practice, $\Psi_\mathrm{RP}$ 
has to be replaced in Eq.~\ref{eq:fourier} by the reconstructed "event plane"
$\Psi_n$ measured from the $n$-th harmonic anisotropy of the event itself. As a
result, the observed flow coefficients $v_n^\textrm{\scriptsize obs}$ have a
magnitude smaller than the real flow coefficients $v_n$. Nevertheless, the true
$v_n$ can be approached by correcting the observed flow parameters
$v_n^\textrm{\scriptsize obs}$ knowing the event plane resolution
$\mathcal{R}_{\Psi_n}$, i.e. the accuracy with which $\Psi_n$ reproduces the
true orientation of the reaction plane $\Psi_\mathrm{RP}$: $v_n =
v_n^\textrm{\scriptsize obs}/\mathcal{R}_{\Psi_n}$. 

\begin{figure}
\centering
\includegraphics[width=0.7\columnwidth]{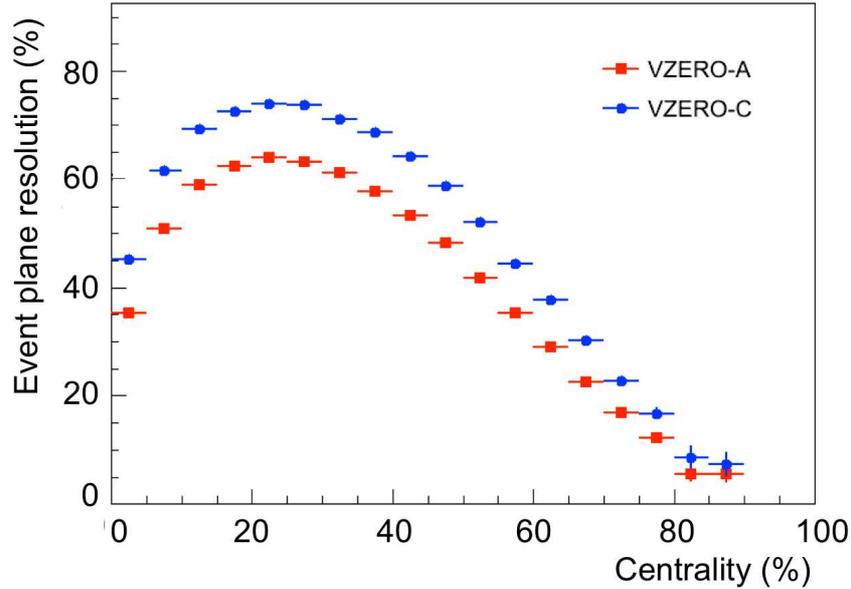}
\caption{Second harmonic event plane resolution $\mathcal{R}_{\Psi_2}$ of
VZERO-A and VZERO-C arrays as 
a function of centrality percentile.}
\label{fig:epResol}
\end{figure}

The resolution can be extracted experimentally using a sub-event correlation
technique~\cite{Poskanzer}. 
Figure~\ref{fig:epResol} shows the strong centrality dependence of the event
plane resolution for the 
second harmonic. The event plane can be best determined for semi-central
collisions where the flow 
effect is largest. The more isotropic the azimuthal event shape is, the more
difficult is the extraction 
of the event plane and the poorer the resolution.

\section{Conclusion}
\label{sec:conclusion}

The VZERO is a detector system which plays a crucial role in ALICE. Over the
first four years of operation at the LHC, the ALICE VZERO system showed an
excellent operational efficiency, which is a necessary requirement for a system
providing the lowest level triggers (L0) for the experiment. In addition, the
VZERO system fulfills several important functions making use of its good time
and charge resolution down to minimum ionizing particles. It is a powerful tool
to reject beam-induced backgrounds and to measure and monitor the LHC
luminosity. Moreover, it is also used to measure global properties of pp and
Pb--Pb collisions such as particle multiplicity, collision centrality and event
plane direction. The reduction of signal due to accumulated radiation doses was
compensated, when necessary, by increasing the high voltage on each PMT. These
adjustments have kept the detector fully operational without degrading its
performance in time and charge resolution, except for some periods of high beam
background during the 2012 run, when it was turned off as a conservation
measure.   

\newenvironment{acknowledgement}{\relax}{\relax}
\begin{acknowledgement}
\section*{Acknowledgements}
The ALICE collaboration would like to thank all its engineers and technicians for their invaluable contributions to the construction of the experiment and the CERN accelerator teams for the outstanding performance of the LHC complex.
\\
The ALICE collaboration acknowledges the following funding agencies for their support in building and
running the ALICE detector:
 \\
State Committee of Science,  World Federation of Scientists (WFS)
and Swiss Fonds Kidagan, Armenia,
 \\
Conselho Nacional de Desenvolvimento Cient\'{\i}fico e Tecnol\'{o}gico (CNPq), Financiadora de Estudos e Projetos (FINEP),
Funda\c{c}\~{a}o de Amparo \`{a} Pesquisa do Estado de S\~{a}o Paulo (FAPESP);
 \\
National Natural Science Foundation of China (NSFC), the Chinese Ministry of Education (CMOE)
and the Ministry of Science and Technology of China (MSTC);
 \\
Ministry of Education and Youth of the Czech Republic;
 \\
Danish Natural Science Research Council, the Carlsberg Foundation and the Danish National Research Foundation;
 \\
The European Research Council under the European Community's Seventh Framework Programme;
 \\
Helsinki Institute of Physics and the Academy of Finland;
 \\
French CNRS-IN2P3, the `Region Pays de Loire', `Region Alsace', `Region Auvergne' and CEA, France;
 \\
German BMBF and the Helmholtz Association;
\\
General Secretariat for Research and Technology, Ministry of
Development, Greece;
\\
Hungarian OTKA and National Office for Research and Technology (NKTH);
 \\
Department of Atomic Energy and Department of Science and Technology of the Government of India;
 \\
Istituto Nazionale di Fisica Nucleare (INFN) and Centro Fermi -
Museo Storico della Fisica e Centro Studi e Ricerche "Enrico
Fermi", Italy;
 \\
MEXT Grant-in-Aid for Specially Promoted Research, Ja\-pan;
 \\
Joint Institute for Nuclear Research, Dubna;
 \\
National Research Foundation of Korea (NRF);
 \\
CONACYT, DGAPA, M\'{e}xico, ALFA-EC and the EPLANET Program
(European Particle Physics Latin American Network)
 \\
Stichting voor Fundamenteel Onderzoek der Materie (FOM) and the Nederlandse Organisatie voor Wetenschappelijk Onderzoek (NWO), Netherlands;
 \\
Research Council of Norway (NFR);
 \\
Polish Ministry of Science and Higher Education;
 \\
National Authority for Scientific Research - NASR (Autoritatea Na\c{t}ional\u{a} pentru Cercetare \c{S}tiin\c{t}ific\u{a} - ANCS);
 \\
Ministry of Education and Science of Russian Federation, Russian
Academy of Sciences, Russian Federal Agency of Atomic Energy,
Russian Federal Agency for Science and Innovations and The Russian
Foundation for Basic Research;
 \\
Ministry of Education of Slovakia;
 \\
Department of Science and Technology, South Africa;
 \\
CIEMAT, EELA, Ministerio de Econom\'{i}a y Competitividad (MINECO) of Spain, Xunta de Galicia (Conseller\'{\i}a de Educaci\'{o}n),
CEA\-DEN, Cubaenerg\'{\i}a, Cuba, and IAEA (International Atomic Energy Agency);
 \\
Swedish Research Council (VR) and Knut $\&$ Alice Wallenberg
Foundation (KAW);
 \\
Ukraine Ministry of Education and Science;
 \\
United Kingdom Science and Technology Facilities Council (STFC);
 \\
The United States Department of Energy, the United States National
Science Foundation, the State of Texas, and the State of Ohio.
\end{acknowledgement}

\bibliographystyle{JHEP}

\newpage
\appendix
\section{The ALICE Collaboration}
\label{app:collab}

\begingroup
\small
\begin{flushleft}
E.~Abbas\Irefn{org36632}\And
B.~Abelev\Irefn{org1234}\And
J.~Adam\Irefn{org1274}\And
D.~Adamov\'{a}\Irefn{org1283}\And
A.M.~Adare\Irefn{org1260}\And
M.M.~Aggarwal\Irefn{org1157}\And
G.~Aglieri~Rinella\Irefn{org1192}\And
M.~Agnello\Irefn{org1313}\textsuperscript{,}\Irefn{org1017688}\And
A.G.~Agocs\Irefn{org1143}\And
A.~Agostinelli\Irefn{org1132}
% ...
\renewcommand\labelenumi{\textsuperscript{\theenumi}~}
\section*{Affiliation notes}
\renewcommand\theenumi{\roman{enumi}}
\begin{Authlist}
\item \Adef{0}Deceased
\item \Adef{M.V.Lomonosov Moscow State University, D.V.Skobeltsyn Institute of Nuclear Physics, Moscow, Russia}Also at: M.V.Lomonosov Moscow State University, D.V.Skobeltsyn Institute of Nuclear Physics, Moscow, Russia
\item \Adef{University of Belgrade, Faculty of Physics and "Vinvca" Institute of Nuclear Sciences, Belgrade, Serbia}Also at: University of Belgrade, Faculty of Physics and "Vin$\rm \breve{c}$a" Institute of Nuclear Sciences, Belgrade, Serbia
\item \Adef{Institute of Theoretical Physics, University of Wroclaw, Wroclaw, Poland}Also at: Institute of Theoretical Physics, University of Wroclaw, Wroclaw, Poland
\end{Authlist}
\section*{Collaboration Institutes}
\renewcommand\theenumi{\arabic{enumi}~}
\begin{Authlist}
\item \Idef{org36632}Academy of Scientific Research and Technology (ASRT), Cairo, Egypt
\item \Idef{org1332}A. I. Alikhanyan National Science Laboratory (Yerevan Physics Institute) Foundation, Yerevan, Armenia
\item \Idef{org1279}Benem\'{e}rita Universidad Aut\'{o}noma de Puebla, Puebla, Mexico
\item \Idef{org1220}Bogolyubov Institute for Theoretical Physics, Kiev, Ukraine
\item \Idef{org20959}Bose Institute, Department of Physics and Centre for Astroparticle Physics and Space Science (CAPSS), Kolkata, India
\item \Idef{org1262}Budker Institute for Nuclear Physics, Novosibirsk, Russia
\item \Idef{org1292}California Polytechnic State University, San Luis Obispo, California, United States
\item \Idef{org1329}Central China Normal University, Wuhan, China
\item \Idef{org14939}Centre de Calcul de l'IN2P3, Villeurbanne, France
\item \Idef{org1197}Centro de Aplicaciones Tecnol\'{o}gicas y Desarrollo Nuclear (CEADEN), Havana, Cuba
\item \Idef{org1242}Centro de Investigaciones Energ\'{e}ticas Medioambientales y Tecnol\'{o}gicas (CIEMAT), Madrid, Spain
\item \Idef{org1244}Centro de Investigaci\'{o}n y de Estudios Avanzados (CINVESTAV), Mexico City and M\'{e}rida, Mexico
\item \Idef{org1335}Centro Fermi - Museo Storico della Fisica e Centro Studi e Ricerche ``Enrico Fermi'', Rome, Italy
\item \Idef{org17347}Chicago State University, Chicago, United States
\item \Idef{org1288}Commissariat \`{a} l'Energie Atomique, IRFU, Saclay, France
\item \Idef{org15782}COMSATS Institute of Information Technology (CIIT), Islamabad, Pakistan
\item \Idef{org1294}Departamento de F\'{\i}sica de Part\'{\i}culas and IGFAE, Universidad de Santiago de Compostela, Santiago de Compostela, Spain
\item \Idef{org1106}Department of Physics Aligarh Muslim University, Aligarh, India
\item \Idef{org1121}Department of Physics and Technology, University of Bergen, Bergen, Norway
\item \Idef{org1162}Department of Physics, Ohio State University, Columbus, Ohio, United States
\item \Idef{org1300}Department of Physics, Sejong University, Seoul, South Korea
\item \Idef{org1268}Department of Physics, University of Oslo, Oslo, Norway
\item \Idef{org1315}Dipartimento di Fisica dell'Universit\`{a} and Sezione INFN, Trieste, Italy
\item \Idef{org1145}Dipartimento di Fisica dell'Universit\`{a} and Sezione INFN, Cagliari, Italy
\item \Idef{org1312}Dipartimento di Fisica dell'Universit\`{a} and Sezione INFN, Turin, Italy
\item \Idef{org1285}Dipartimento di Fisica dell'Universit\`{a} `La Sapienza' and Sezione INFN, Rome, Italy
\item \Idef{org1154}Dipartimento di Fisica e Astronomia dell'Universit\`{a} and Sezione INFN, Catania, Italy
\item \Idef{org1132}Dipartimento di Fisica e Astronomia dell'Universit\`{a} and Sezione INFN, Bologna, Italy
\item \Idef{org1270}Dipartimento di Fisica e Astronomia dell'Universit\`{a} and Sezione INFN, Padova, Italy
\item \Idef{org1290}Dipartimento di Fisica `E.R.~Caianiello' dell'Universit\`{a} and Gruppo Collegato INFN, Salerno, Italy
\item \Idef{org1103}Dipartimento di Scienze e Innovazione Tecnologica dell'Universit\`{a} del Piemonte Orientale and Gruppo Collegato INFN, Alessandria, Italy
\item \Idef{org1114}Dipartimento Interateneo di Fisica `M.~Merlin' and Sezione INFN, Bari, Italy
\item \Idef{org1237}Division of Experimental High Energy Physics, University of Lund, Lund, Sweden
\item \Idef{org1192}European Organization for Nuclear Research (CERN), Geneva, Switzerland
\item \Idef{org1227}Fachhochschule K\"{o}ln, K\"{o}ln, Germany
\item \Idef{org1122}Faculty of Engineering, Bergen University College, Bergen, Norway
\item \Idef{org1136}Faculty of Mathematics, Physics and Informatics, Comenius University, Bratislava, Slovakia
\item \Idef{org1274}Faculty of Nuclear Sciences and Physical Engineering, Czech Technical University in Prague, Prague, Czech Republic
\item \Idef{org1229}Faculty of Science, P.J.~\v{S}af\'{a}rik University, Ko\v{s}ice, Slovakia
\item \Idef{org1184}Frankfurt Institute for Advanced Studies, Johann Wolfgang Goethe-Universit\"{a}t Frankfurt, Frankfurt, Germany
\item \Idef{org1215}Gangneung-Wonju National University, Gangneung, South Korea
\item \Idef{org20958}Gauhati University, Department of Physics, Guwahati, India
\item \Idef{org1212}Helsinki Institute of Physics (HIP) and University of Jyv\"{a}skyl\"{a}, Jyv\"{a}skyl\"{a}, Finland
\item \Idef{org1203}Hiroshima University, Hiroshima, Japan
\item \Idef{org1254}Indian Institute of Technology Bombay (IIT), Mumbai, India
\item \Idef{org36378}Indian Institute of Technology Indore, Indore, India (IITI)
\item \Idef{org1266}Institut de Physique Nucl\'{e}aire d'Orsay (IPNO), Universit\'{e} Paris-Sud, CNRS-IN2P3, Orsay, France
\item \Idef{org1277}Institute for High Energy Physics, Protvino, Russia
\item \Idef{org1249}Institute for Nuclear Research, Academy of Sciences, Moscow, Russia
\item \Idef{org1320}Nikhef, National Institute for Subatomic Physics and Institute for Subatomic Physics of Utrecht University, Utrecht, Netherlands
\item \Idef{org1250}Institute for Theoretical and Experimental Physics, Moscow, Russia
\item \Idef{org1230}Institute of Experimental Physics, Slovak Academy of Sciences, Ko\v{s}ice, Slovakia
\item \Idef{org1127}Institute of Physics, Bhubaneswar, India
\item \Idef{org1275}Institute of Physics, Academy of Sciences of the Czech Republic, Prague, Czech Republic
\item \Idef{org1139}Institute of Space Sciences (ISS), Bucharest, Romania
\item \Idef{org27399}Institut f\"{u}r Informatik, Johann Wolfgang Goethe-Universit\"{a}t Frankfurt, Frankfurt, Germany
\item \Idef{org1185}Institut f\"{u}r Kernphysik, Johann Wolfgang Goethe-Universit\"{a}t Frankfurt, Frankfurt, Germany
\item \Idef{org1177}Institut f\"{u}r Kernphysik, Technische Universit\"{a}t Darmstadt, Darmstadt, Germany
\item \Idef{org1256}Institut f\"{u}r Kernphysik, Westf\"{a}lische Wilhelms-Universit\"{a}t M\"{u}nster, M\"{u}nster, Germany
\item \Idef{org1246}Instituto de Ciencias Nucleares, Universidad Nacional Aut\'{o}noma de M\'{e}xico, Mexico City, Mexico
\item \Idef{org1247}Instituto de F\'{\i}sica, Universidad Nacional Aut\'{o}noma de M\'{e}xico, Mexico City, Mexico
\item \Idef{org1308}Institut Pluridisciplinaire Hubert Curien (IPHC), Universit\'{e} de Strasbourg, CNRS-IN2P3, Strasbourg, France
\item \Idef{org1182}Joint Institute for Nuclear Research (JINR), Dubna, Russia
\item \Idef{org1199}Kirchhoff-Institut f\"{u}r Physik, Ruprecht-Karls-Universit\"{a}t Heidelberg, Heidelberg, Germany
\item \Idef{org20954}Korea Institute of Science and Technology Information, Daejeon, South Korea
\item \Idef{org1017642}KTO Karatay University, Konya, Turkey
\item \Idef{org1160}Laboratoire de Physique Corpusculaire (LPC), Clermont Universit\'{e}, Universit\'{e} Blaise Pascal, CNRS--IN2P3, Clermont-Ferrand, France
\item \Idef{org1194}Laboratoire de Physique Subatomique et de Cosmologie (LPSC), Universit\'{e} Joseph Fourier, CNRS-IN2P3, Institut Polytechnique de Grenoble, Grenoble, France
\item \Idef{org1187}Laboratori Nazionali di Frascati, INFN, Frascati, Italy
\item \Idef{org1232}Laboratori Nazionali di Legnaro, INFN, Legnaro, Italy
\item \Idef{org1125}Lawrence Berkeley National Laboratory, Berkeley, California, United States
\item \Idef{org1234}Lawrence Livermore National Laboratory, Livermore, California, United States
\item \Idef{org1251}Moscow Engineering Physics Institute, Moscow, Russia
\item \Idef{org1322}National Centre for Nuclear Studies, Warsaw, Poland
\item \Idef{org1140}National Institute for Physics and Nuclear Engineering, Bucharest, Romania
\item \Idef{org1017626}National Institute of Science Education and Research, Bhubaneswar, India
\item \Idef{org1165}Niels Bohr Institute, University of Copenhagen, Copenhagen, Denmark
\item \Idef{org1109}Nikhef, National Institute for Subatomic Physics, Amsterdam, Netherlands
\item \Idef{org1283}Nuclear Physics Institute, Academy of Sciences of the Czech Republic, \v{R}e\v{z} u Prahy, Czech Republic
\item \Idef{org1264}Oak Ridge National Laboratory, Oak Ridge, Tennessee, United States
\item \Idef{org1189}Petersburg Nuclear Physics Institute, Gatchina, Russia
\item \Idef{org1170}Physics Department, Creighton University, Omaha, Nebraska, United States
\item \Idef{org1157}Physics Department, Panjab University, Chandigarh, India
\item \Idef{org1112}Physics Department, University of Athens, Athens, Greece
\item \Idef{org1152}Physics Department, University of Cape Town and  iThemba LABS, National Research Foundation, Somerset West, South Africa
\item \Idef{org1209}Physics Department, University of Jammu, Jammu, India
\item \Idef{org1207}Physics Department, University of Rajasthan, Jaipur, India
\item \Idef{org1200}Physikalisches Institut, Ruprecht-Karls-Universit\"{a}t Heidelberg, Heidelberg, Germany
\item \Idef{org1017688}Politecnico di Torino, Turin, Italy
\item \Idef{org1325}Purdue University, West Lafayette, Indiana, United States
\item \Idef{org1281}Pusan National University, Pusan, South Korea
\item \Idef{org1176}Research Division and ExtreMe Matter Institute EMMI, GSI Helmholtzzentrum f\"ur Schwerionenforschung, Darmstadt, Germany
\item \Idef{org1334}Rudjer Bo\v{s}kovi\'{c} Institute, Zagreb, Croatia
\item \Idef{org1298}Russian Federal Nuclear Center (VNIIEF), Sarov, Russia
\item \Idef{org1252}Russian Research Centre Kurchatov Institute, Moscow, Russia
\item \Idef{org1224}Saha Institute of Nuclear Physics, Kolkata, India
\item \Idef{org1130}School of Physics and Astronomy, University of Birmingham, Birmingham, United Kingdom
\item \Idef{org1338}Secci\'{o}n F\'{\i}sica, Departamento de Ciencias, Pontificia Universidad Cat\'{o}lica del Per\'{u}, Lima, Peru
\item \Idef{org1155}Sezione INFN, Catania, Italy
\item \Idef{org1313}Sezione INFN, Turin, Italy
\item \Idef{org1271}Sezione INFN, Padova, Italy
\item \Idef{org1133}Sezione INFN, Bologna, Italy
\item \Idef{org1146}Sezione INFN, Cagliari, Italy
\item \Idef{org1316}Sezione INFN, Trieste, Italy
\item \Idef{org1115}Sezione INFN, Bari, Italy
\item \Idef{org1286}Sezione INFN, Rome, Italy
\item \Idef{org36377}Nuclear Physics Group, STFC Daresbury Laboratory, Daresbury, United Kingdom
\item \Idef{org1258}SUBATECH, Ecole des Mines de Nantes, Universit\'{e} de Nantes, CNRS-IN2P3, Nantes, France
\item \Idef{org35706}Suranaree University of Technology, Nakhon Ratchasima, Thailand
\item \Idef{org1304}Technical University of Split FESB, Split, Croatia
\item \Idef{org1017659}Technische Universit\"{a}t M\"{u}nchen, Munich, Germany
\item \Idef{org1168}The Henryk Niewodniczanski Institute of Nuclear Physics, Polish Academy of Sciences, Cracow, Poland
\item \Idef{org17361}The University of Texas at Austin, Physics Department, Austin, TX, United States
\item \Idef{org1173}Universidad Aut\'{o}noma de Sinaloa, Culiac\'{a}n, Mexico
\item \Idef{org1296}Universidade de S\~{a}o Paulo (USP), S\~{a}o Paulo, Brazil
\item \Idef{org1149}Universidade Estadual de Campinas (UNICAMP), Campinas, Brazil
\item \Idef{org1239}Universit\'{e} de Lyon, Universit\'{e} Lyon 1, CNRS/IN2P3, IPN-Lyon, Villeurbanne, France
\item \Idef{org1205}University of Houston, Houston, Texas, United States
\item \Idef{org20371}University of Technology and Austrian Academy of Sciences, Vienna, Austria
\item \Idef{org1222}University of Tennessee, Knoxville, Tennessee, United States
\item \Idef{org1310}University of Tokyo, Tokyo, Japan
\item \Idef{org1318}University of Tsukuba, Tsukuba, Japan
\item \Idef{org21360}Eberhard Karls Universit\"{a}t T\"{u}bingen, T\"{u}bingen, Germany
\item \Idef{org1225}Variable Energy Cyclotron Centre, Kolkata, India
\item \Idef{org1017687}Vestfold University College, Tonsberg, Norway
\item \Idef{org1306}V.~Fock Institute for Physics, St. Petersburg State University, St. Petersburg, Russia
\item \Idef{org1323}Warsaw University of Technology, Warsaw, Poland
\item \Idef{org1179}Wayne State University, Detroit, Michigan, United States
\item \Idef{org1143}Wigner Research Centre for Physics, Hungarian Academy of Sciences, Budapest, Hungary
\item \Idef{org1260}Yale University, New Haven, Connecticut, United States
\item \Idef{org15649}Yildiz Technical University, Istanbul, Turkey
\item \Idef{org1301}Yonsei University, Seoul, South Korea
\item \Idef{org1327}Zentrum f\"{u}r Technologietransfer und Telekommunikation (ZTT), Fachhochschule Worms, Worms, Germany
\end{Authlist}
\endgroup

%
%
\end{document}